\documentclass[12pt]{article}

\usepackage{amsfonts,amssymb,amsmath}

\newcommand{\bq}{{\mathbf q}}

\newcommand{\ra}{{\rightarrow}}

\newcommand{\rank}{{\rm rank}}

\newcommand{\eps}{\epsilon}

\begin{document}

\author{Anton Kapustin\\ {\it California Institute of Technology}}

\title{Remarks on nonrelativistic Goldstone bosons}

\begin{titlepage}

\maketitle

\abstract{We discuss excitations in nonrelativistic field theories with spontaneous breaking of a continuous global symmetry. It is known that in such systems there are two types of Goldstone bosons (Type A and Type B) whose dispersion law is generically linear or quadratic, respectively. We show that Type B Goldstone bosons may have gapped partners which we call almost-Goldstone bosons. With some nondegeneracy assumption about the low-energy effective action, the total number of Goldstone and almost-Goldstone bosons adds up to the number of broken symmetry generators. We propose that deviations of the dispersion law of Goldstone bosons from linearity at small momenta may serve as a signature of small breaking of time-reversal symmetry.}

\end{titlepage}

\section{Introduction}

In its most general form, the Goldstone theorem says that in a translationally-invariant system spontaneous breaking of a continuous symmetry $G$ produces gapless bosonic excitations, i.e. excitations whose energy vanishes as the spatial momentum $\bq$ goes to zero. These gapless excitations are called Goldstone bosons. In the relativistic case, one can show that there is one Goldstone boson for every generator of the symmetry which does not annihilate the vacuum. It was noted by Nielsen and Chadha \cite{NC} that if one abandons Lorenz invariance, the situation is more complex because there exist two different types of Goldstone bosons which they referred to as Type I and Type II Goldstones. By definition, a Type I Goldstone boson has energy which scales as an odd power of momentum as $\bq\ra 0$, while a Type II Goldstone boson has energy which scales as an even power of momentum. Suppose the number of Type I and Type II Goldstones is $n_I$ and $n_{II}$ respectively, and the number  of broken symmetry generators is $N$. Then there is an inequality \cite{NC}:
$$
n_I+2n_{II}\geq N.
$$
Later Leutwyler \cite{L} and Nambu \cite{Nambu} pointed out that the presence of Type II Goldstone bosons is associated with a nonzero charge density for a charge which is a commutator of two broken charges. For a review of these and other related works see \cite{Brauner}. More recently Watanabe and Brauner \cite{WB} conjectured that the numbers of Type I and Type II Goldstone bosons satisfy a relation
\begin{equation}\label{WBeq}
n_I+n_{II}=N-\frac12 \rank\, B
\end{equation}
where $B$ is an $N\times N$ matrix which encodes the commutators of broken charges:
$$
B_{ij}=\lim_{V\ra\infty}\frac{-i}{V}[Q_i,Q_j].
$$
Here $V$ is the spatial volume and $Q_i$, $i=1,\ldots,N$, are the broken generators.
Watanabe and Brauner proved that the left-hand side of eq.~(\ref{WBeq}) is greater or equal than the right-hand side. It is also known that when $B=0$ eq.~(\ref{WBeq}) holds true \cite{Schaferetal}. 

Very recently Watanabe and Murayama \cite{WM} proved the Watanabe-Brauner conjecture using the effective action approach. More precisely, they define Type A and Type B Goldstons, which are closely related to Type I and Type II Goldstones, and show that their numbers are given by
\begin{equation}\label{WMeq}
n_A=N-\rank\, B,\quad n_B=\frac12 \rank\, B.
\end{equation}
Eq. (\ref{WBeq}) is a consequence of these more precise counting formulas.

In this note we refine the analysis of \cite{WM} and show that apart from true Goldstone bosons the effective action considered in \cite{WM} describes gapped excitations which we call almost-Goldstone bosons. We determine their number and show that if the target-space metric is nondegenerate, it is equal to $\frac12\rank\, B$, so that the total number of Goldstone and almost-Goldstone bosons is $N$. We explain a mechanism by which two Type A Goldstone bosons may pair up into a Type B Goldstone boson and an almost-Goldstone boson. The number of such pairs is precisely $\rank\, B$. This gives a simple and intuitive explanation of the counting rules eq.~(\ref{WMeq}). 

In Section 2 we analyze the effective action for the order parameter and compute the number of Goldstone and almost-Goldstone bosons. We follow closely \cite{WM}, but remove some unnecessary assumptions about the form of the action. In Section 3 we give some examples and discuss our results. In particular, we propose that deviations from linearity in the dispersion law of Goldstone bosons at small momenta serve as a signature of a small breaking of time-reversal symmetry.

I would like to thank Ira Rothstein for discussions and Hiroshi Ooguri for drawing my attention to Ref.~\cite{WM}. This work was supported in part by the DOE grant DE-FG02-92ER40701.

\section{Goldstone and almost-Goldstone bosons}

Suppose the symmetry group $G$ (which we assume for now to be internal symmetry, i.e. it does not act on time or spatial coordinates) is spontaneously broken down to a subgroup $H$, by which we mean that there is an order parameter taking values in $G/H$. Our basic assumption is that the low-energy theory can be described by an action for a field $\phi$ taking values in $G/H$. One further assumption is that the action contains terms only of first or second order in time derivatives. Thus it has the form
$$
S=\int dt d^nx \left( \frac12 G_{ij}(\phi) \partial_t\phi^i \partial_t\phi^j+A_i(\phi,\nabla)\partial_t\phi^i-W(\phi,\nabla)\right).
$$
By $A_i(\phi,\nabla), etc.$ we mean some function of $\phi$ and its spatial derivatives. We do not assume rotational invariance. For simplicity we assumed that the term quadratic in time derivatives does not depend on spatial derivatives, although this is not really necessary. The above action is slightly more general than that considered in \cite{WM}, where $A_i$ was assumed to depend on the field $\phi$, but not on its spatial derivatives.

If the target space metric $G_{ij}$ is  nondegenerate, we can rewrite the action in the first-order form as follows:
$$
S=\int dt d^nx \left[\left(p_i+A_i(\phi,\nabla)\right)\partial_t\phi^i-\frac12 G^{ij}(\phi)p_i p_j -W(\phi,\nabla)\right],
$$
where $G^{ij}$ is the inverse of $G_{ij}$.
We can now redefine momenta
$$
p_i\mapsto p_i-A_i(\phi,\nabla),
$$
and bring the action to the standard form 
$$
S=\int dt d^n x \left[p_i\partial_t\phi^i-H(p,\phi,\nabla)\right]
$$
with the Hamiltonian density
$$
H(p,\phi,\nabla\phi)=\frac12 G^{ij}(p_i-A_i)(p_j-A_j)+W(\phi,\nabla).
$$

Let us examine the physics described by this action. We pick any constant vacuum configuration $\phi^i=\phi^i_0$ and expand the Hamiltonian to quadratic order in the fluctuations. Since the target space $G/H$ is a homogeneous $G$-space, and by assumption the action is $G$-invariant, the physics is independent of the choice of $\phi^i_0$, and without loss of generality we may set it to zero. The quadratic part of the Hamiltonian density has the form
$$
H_0=\frac12 g^{ij}_0 (p_i-B_{ik}(-i\nabla)\phi^k)(p_j-B_{jl}(-i\nabla)\phi^l)+\frac12 \phi^i\Omega^2_{ij}(-i\nabla)\phi^j.
$$
Here $g_0^{ij}=G^{ij}(0)$, $B_{ik}(-i\nabla)$ is a matrix whose entries are polynomials in the spatial derivatives such that $B_{ik}(-i\nabla)\phi^k$ is a linearization of $A_i(\phi,\nabla)$, and $\Omega^2_{ij}(-i\nabla)$ is similarly a matrix of spatial differential operators such that $\frac12 \phi^i\Omega^2_{ij}(-i\nabla)\phi^j$ is the leading (quadratic) part in the expansion of $W(\phi,\nabla)$. Since it is assumed that $\phi^i=0$ is a solution of the equations of motion, there are no linear terms in the expansion of $W$. We also absorbed possible constant terms in the expansion of $A_i(\phi,\nabla)$ into a shift of $p_i$. 

Fourier-expanding both $p_i$ and $\phi^i$, we see that  the Hamiltonian for Fourier modes with momentum $\bq$ is almost identical to the Hamiltonian of an anisotropic $N$-dimensional harmonic oscillator in a magnetic field. The only difference is that the canonical coordinates and momenta are complex and subject to the reality constraint $p_i(\bq)^*=p_i(-\bq)$ and $\phi^i(\bq)^*=\phi^i(-\bq)$. We also have the relations $B_{ik}^*(\bq)=B_{ki}(-\bq)$ and $\Omega^2_{ij}(\bq)=\Omega^2_{ij}(-\bq)$. In addition, $\Omega^2_{ij}(\bq)$ is a positive-definite Hermitian matrix.  One can diagonalize the Hamiltonian in the usual way by introducing creation and annihilation operators. This is facilitated by working in a coordinate system where $g^{ij}_0=\delta^{ij}$ and by shifting
$$
p_i(\bq)\mapsto p_i(\bq)+\frac12 (B_{ij}(\bq)+B_{ji}(-\bq))\phi^j(\bq).
$$
This redefinition does not affect the commutation relations, and its only effect is to replace $B_{ij}(\bq)$ with its ``antisymmetrized'' part $\frac12(B_{ij}(\bq)-B_{ji}(-\bq))$. In other words, we may assume that the matrix function $B_{ij}$ satisfies $B_{ij}(\bq)=-B_{ji}(-\bq)$. Together with the reality condition, this means that $B_{ij}(\bq)$ is an anti-Hermitian matrix.

In terms of the usual creation and annihilation operators the normal-ordered Hamiltonian  takes the form
$$
\int d^n\bq\ a_i(\bq)^\dagger  \left( {\sqrt {\Omega^2(\bq)+B(\bq)^\dagger B(\bq)}}+i B(\bq)\right)_{ij}a_j(\bq).
$$
Here it is understood that we take the positive square root of the positive Hermitian matrix $\Omega^2+B^\dagger B$. 
This Hamiltonian describes $N$ species of bosonic particles. The energies of one-particle excitations with momentum $\bq$ are the eigenvalues of the Hermitian matrix
\begin{equation}\label{M}
M(\bq)=iB(\bq)+\sqrt {\Omega^2(\bq)+B(\bq)^\dagger B(\bq)}
\end{equation}

We can now determine the number of gapless excitations arising from the fluctuations of the order parameter, i.e. the number of Goldstone bosons. Consider the limit $\bq\ra 0$. In this limit the stiffness matrix $\Omega^2(\bq)$ necessarily goes to zero, since otherwise $\phi^i=const$ would not be a solution of the classical equations of motion. The energies of the $\bq=0$ excitations are therefore the eigenvalues of the matrix
$$
M(0)=iB(0)+\sqrt {B(0)^\dagger B(0)}
$$
The matrix $B(0)$ is a real skew-symmetric matrix.
We can use an orthogonal transformation to bring it to the standard block-diagonal form
$$
B(0)=\begin{pmatrix} 0 & b_1 &     &   &  &  &  \\
                                 -b_1    &    0    &   &  &  &  \\
                                     &        &  \hdotsfor{2}   &  &   &  \\
                                     &        &\hdotsfor{2} &  &   &   \\
                                     &        &     &   & 0 & b_{[\frac{N}{2}]} & \\
                                     &        &     &   & -b_{[\frac{N}{2}]} & 0 &  \\
                                     &        &     &   &                  &    &  [0]\end{pmatrix}
                              $$
Here the brackets around zero in the last row indicate that this diagonal element is present only for odd $N$. Therefore the eigenvalues of the matrix $M(0)$ (i.e. the energies of one-particle excitations with zero momentum) are $2|b_1|,\ldots,2|b_{[N/2]}|,0,\ldots,0$, where the number of zeros is $N-[N/2]=[(N+1)/2]$. Thus the number of Goldstone bosons is $N-\frac12\rank\, B(0)$ \cite{WM}. It can range from $[(N+1)/2]$ to $N$. The remaining $\frac12\rank\, B(0)$ excitations are gapped. We will refer to the gapped excitations as almost-Golstone bosons, since they also arise from small fluctuations of the order parameter. However, it should be kept in mind that if the energy gap for almost-Goldstone bosons is comparable to the masses of other excitations which we integrated out to arrive at our effective action, then the effective action is not valid at these energy scales, and the almost-Goldstone bosons will mix with other excitations. Thus the notion of an almost-Goldstone is well-defined only if nonzero eigenvalues of $B(0)$ are much smaller than the UV energy cutoff.

As for true Goldstone bosons, they can be further classified as being of Type A or Type B depending on whether for $\bq=0$ their internal-space polarizations are in the kernel of $B(0)$ or not \cite{WM}. The number of Type A Goldstone bosons is $\dim \ker B(0)=N-\rank\, B(0)$. The number of Type B Goldstone bosons is $\frac12 \rank\, B(0)$, because each of the Type B Goldstone bosons has an almost-Goldstone partner. The energy gap for the $\ell^{\rm th}$ almost-Goldstone boson is $2|b_\ell|$, where $b_\ell$ is a nonzero eigenvalue of $B(0)$ and $\ell$ runs over $\frac12 \rank\, B(0)$ values. The total number of independent excitations adds up to $N$, of course.
 
Typically, rotational invariance or spatial parity invariance dictates that the stiffness matrix $\Omega^2(\bq)$ is of order $\bq^2$, while the eigenvalues $\pm b_\ell(\bq)$ of the matrix $i B(\bq)$ for small $\bq$ behave as $b_\ell(\bq)=b_\ell(0)+O(\bq^2)$ (some exceptions will be noted below). Barring accidental cancellations,  this means that Type B Goldstone bosons have quadratic dispersion law, while Type  A Goldstone bosons have linear dispersion law.  Thus this classification scheme is consistent with that of \cite{NC}. The effective action approach has the advantage that the counting rules for Type A and Type B Goldstones hold even when the stiffness $\Omega^2(\bq)$ is softer than usual due to accidental cancellations or fine-tuning. 

So far we have been assuming that the metric $G_{ij}(\phi)$ is nondegenerate. The general case is not very different. Consider first the opposite extreme, $G_{ij}=0$. In this case we are dealing with an action which is of first order in time derivatives, and for this action to define a sensible theory $B_{ij}(\bq)$ has to be nondegenerate for all $\bq$, including $\bq=0$. This means, of course, that $N$ has to be even, that $B_{ij}(\bq)$ plays the role of a symplectic structure on the space of fluctuations with momentum $\bq$, and that half of the fluctuations $\phi^i(\bq)$ should be regarded as canonically conjugate to the other half \cite{WM}. This immediately implies that there are $N/2$ independent excitations. For each $\bq$ the Hamiltonian can be thought of as describing a zero-mass  charged particle in a harmonic potential and magnetic field (in $N$ dimensions). The zero-mass limit means that only the lowest Landau level survives.

One can think of this case as arising from the case of nondegenerate $G_{ij}$ in the limit $G_{ij}\ra 0$. Equivalently, if one works in the coordinate system where $g_0^{ij}=\delta^{ij}$, one can rescale $B\mapsto \lambda B$, $\Omega^2\mapsto\lambda\Omega^2$ and take the limit $\lambda\ra\infty$. In this limit half of the eigenvalues of the matrix $M(\bq)$ (the ones corresponding to the almost-Goldstone bosons) go to infinity, while the other half (corresponding to Type B Goldstone bosons) remain finite. Thus we get $N/2$ Goldstones of Type B and no other excitations. This agrees with the analysis of \cite{WM}.

The most general case is now clear. If $G_{ij}$ is degenerate (i.e. positive semi-definite rather than positive-definite), the matrix $B_{ij}(\bq)$ must be nondegenerate when restricted to the zero subspace $\ker G$ of $G_{ij}$, for the action to describe a sensible theory. This means in particular that $\ker G$ is even-dimensional. Fluctuations in the zero subspace of $G_{ij}$ are pairwise canonically conjugate, therefore the total number of independent one-particle excitations is $N-\frac12 \dim\ker G$. Some of these are gapless (true Goldstone bosons), while the rest are gapped (almost-Goldstone bosons). The count of Goldstone bosons is independent of the form of $G$ and depends only on $B(0)$. Namely, we have $N-\frac12 \rank\, B(0)$ Goldstone bosons, out of which $\frac12 \rank\, B(0)$ are Type B and $N-\rank\, B(0)$ are Type A. The number of almost-Goldstone bosons is 
$$
\frac12 (\rank\, B(0)-\dim\ker G).
$$

\section{Discussion and examples}

The derivation of the Goldstone boson count just presented clarifies the relationship between Type A and Type B Goldstone bosons. Suppose we start with an action which does not contain terms with only a single time derivative. Such an action describes $N$ Type A Goldstone bosons. If we perturb the theory by adding a term with a single time derivative, some Goldstone bosons (namely, the ones which lie in nonzero eigenspaces of $B(0)$) are paired up, and each pair gives rise to a single Type B Goldstone boson and a single almost-Goldstone boson. This is why the number of true Goldstone bosons is now decreased to $N-\frac12\rank\, B(0)$. This bears some resemblance to the Higgs effect, where Goldstone bosons become longitudinal polarizations of massive gauge bosons. However, there is a crucial distinction (apart from the fact that no gauge interactions are involved in our case): the Higgs effect is a classical phenomenon, while the emergence of the almost-Goldstone bosons is a quantum effect. This is quite clear from the above derivation, which relies on the existence of Landau levels for a particle in a magnetic field. 

This viewpoint might suggest that in the nonrelativistic case Type B Goldstones are generic, while Type A Goldstones require fine-tuning. This is not so, however, because terms of first order in time derivatives are odd under time reversal, while the rest of the action is even. We assumed here that the order parameter does not transform under time reversal. If this naive time-reversal transformation is a symmetry of the microscopic theory, we must have $B=0$, and Type B Goldstone bosons are forbidden. If the naive time-reversal is not a symmetry, the system might still possess time-reversal symmetry under which the order parameter transforms nontrivially. But in this case there no symmetry reason for $B(0)$ to vanish, and Type B Goldstone bosons are generic. Such is the case of the Heseinberg ferromagnet, where the microscopic theory is time-reversal invariant, but the order parameter (magnetization) is odd under time-reversal. Hence we expect that magnons in the ferromagnetic phase (Goldstone bosons arising from $SO(3)$ breaking down to $SO(2)$) are of Type B. On the other hand, in an antiferromagnet the order parameter is even under time-reversal, so magnons are Type A Goldstones.

If the naive time-reversal symmetry is only slightly broken, then the splitting between a Type B Goldstone boson and its partner almost-Goldstone boson is small, and moreover for moderately large momenta the dispersion law for both is essentially linear, i.e. they are approximately of Type A. The deviations from linearity will be observed only for very small momenta, where the dispersion law for the Goldstone boson is $\eps(\bq)=K_{\alpha\beta} \bq^\alpha\bq^\beta+\ldots$, while for the almost-Goldstone boson it is $\eps(\bq)=2b(0)+K_{\alpha\beta} \bq^\alpha\bq^\beta+\ldots$. One can look for such deviations from linearity at small momenta as a signature of a small breaking of time-reversal symmetry.

There may be other symmetry considerations forbidding Type B Goldstone bosons. Consider a phase where $G\times G$ is spontaneously broken down to the diagonal subgroup $G$. Such is the case for the B-phase of superfluid helium-3, for example, where $G=SO(3)$. If $G$ is compact semi-simple, then no Type B Goldstone bosons are allowed. Indeed, the order parameter takes values in $G$, and all the quantities appearing in the effective action must be invariant with respect to both left and right $G$-action. In particular, the 2-form $B_{ij}(0)$ must be invariant with respect to both left and right $G$-action. It is easy to show that the only such 2-form is $0$. 

On the other hand, consider a phase where a compact semi-simple $G$ is broken down to nothing. The order parameter again takes values in $G$, but now the effective action must be invariant only with respect to the left $G$-action. There are plenty of left-invariant 2-forms on $G$ (just take the wedge product of any two left-invariant 1-forms), so unless time-reversal considerations forbid terms with only a single time-derivatives, Type B Goldstones will be generic. (If $\dim G$ is odd-dimensional, there will be at least one Type A Goldstone boson, since an $N\times N$ skew-symmetric matrix cannot have rank $N$ if $N$ is odd). 

To conclude this section, let us comment on the somewhat peculiar case of one spatial dimension. There can be no Goldstone bosons in one spatial dimension, nevertheless  one may consider actions of the same form as in higher dimensions, and many of the above considerations still apply. Consider for example an action for a single scalar field $\phi$ of the form
$$
S=\int dt dx (\partial_x\phi\partial_t\phi-\partial_x^m\phi \partial_x^m\phi).
$$
For $m=1$ this action describes a chiral boson. The matrix $B(\bq)=i\bq$ is of size $1\times 1$, and so is $\Omega^2(\bq)=\bq^{2M}$. The metric $G$ vanishes identically, so formally the action describes $\frac12 \rank\, B(\bq)=1/2$ bosonic degrees of freedom. What this really means is that the action describes bosonic particles which are right-moving (have $\bq>0$) and have the dispersion law $\eps(\bq)=\bq^m$. In higher dimensions we would reject such an action because the matrix $B(0)$ vanishes identically, so the zero mode of $\phi$ does not have a conjugate momentum. However, in one spatial dimension we typically do not regard the zero mode of $\phi$ as observable: all allowed observables must be invariant under $\phi\mapsto \phi+{\rm const}$. Thus the action gives rise to a sensible theory ``of Type A'', even though the particle it describes cannot be thought of as a Goldstone boson. 

We have started our discussion by stating that $G$ is an internal symmetry. However, this was not really necessary: everything we said applies to any continuous symmetry which does not involve the time coordinate. Since we also assumed translational invariance, in practice this means that $G$ may contain both internal symmetries and rotations.


\end{document}